\documentstyle[12pt]{article}
\normalbaselines
\begin{document}
\pagestyle{empty}

\vbox {\vspace{6mm}}

\begin{center}
{ \bf SQUEEZED STATES AND PARTICLE PRODUCTION\\[2mm]
IN HIGH ENERGY COLLISIONS}\\[7mm]
Bindu A. Bambah\\
{\it School of Physics\\ Univ. Of Hyderabad
Hyderabad-500134,India}\\[5mm]
\end{center}

\vspace{2mm}
\begin{abstract}
Using the `quantum optical approach'  we
propose a model of  multiplicity  distributions  in  high  energy
collisions based on squeezed coherent states.
 We show that the k-mode squeezed coherent state is the most general
one in describing hadronic mulitiplicity distributions in particle collision
processes, describing not only $p \overline p$ collisions but 
$e^{+}e^{-}$, $\nu p$ and diffractive collisions as well.
 The reason for this phenomenological
fit has been gained by working  out a microscopic theory in
which  the  squeezed  coherent  sources  arise  naturally  if  one 
considers the Lorentz  squeezing  of  hadrons  and  works  in  the
covariant phase space formalism .
\end{abstract}

\section{ INTRODUCTION}
Although   Quantum
Chromodynamics is widely believed to  be  the  theory  of  Strong 
Interactions, very few experimental results support this claim.
In particular the behaviour of QCD at small momentum  transfer i.e
low  energies  is   not   understood.
 This  lack  of understanding
reflects itself in the  fact  that  particle  production  in  high 
energy collisions cannot be explained within QCD.
 Given the absence of a detailed dynamical theory of  strong
interactions , one can adopt a  statistical  outlook  and  try  to 
forecast macroscopic behaviour of  a  strongly  interacting  system 
given   only   partial   information    about    their    internal 
states.Experimental information about hadronisation in high energy 
collisions comes from the observation of jets of hadrons  and  the 
distributions of the final state particles.
 By using  analogies
with quantum optical systems one can  get  information  about  the 
types of sources( Chaotic, coherent, etc) that are responsible for 
hadronic emission. Also, by using adapting another quantum optical 
effect such as the Hanbury-Brown Twiss effect one  can  study  the 
size and lifetime of the emitting  region.  This  information  can 
then be used to put restrictions on the microscopic theory pursued 
from the quark-parton end \cite{car}.
  
The experimental quantities amenable to the quantum statistical approach are:
 the multiplicity Distribution of final state particles (PIONS)given by
\begin{equation}P_{n}=\frac{\sigma_{n}}{\sigma_{inel}}\end{equation}
 where
$\sigma_{n} $ n-pion cross-section, 
the number of particles produced per unit rapidity $dN/dy$ , where 
$ y=ln(\frac{E+p_{L}}{E-p_{L}}) $ is the rapidity which plays the role of time 
in pion counting experiments, the moments of $P_{n}$
and the two pion correlations which are analogous to Hanbury Brown Twiss 
effect for pions in rapidity space.

In particular, the quantum optical models are
 based on the assumption that multiparticle production takes place in two 
 stages. In  the initial stage formation of an excited system (fireball) which 
 consists of a number of well defined phase space cells  or 'sources' which 
 then hadronize independently. In these models an ansatz is made about the 
 statistical nature of these sources and the resulting multiplicity 
 distributions are compared with data \cite{fowl}, \cite{meng}.
Table 1. gives the comparison of various quantum optical models.

\begin{table}[h]
\caption{ Comparison of Quantum Optical Models of Multiplicity Distributions}
\begin{tabular}{||l|l|l|l|}     \hline
 Nature Of Source& Density matrix& Probability &Two pion \\ \hline
 One Source&    (Coherent State Rep)&Distribution&Correlations\\ \hline
$P(\alpha)=\frac{1}{\pi n}exp^{(-|\alpha|^{2}/\overline{n})}$&
$\rho_{nm}=\frac{\overline{n}^{n}}{(1+\overline{n})^{n+1}}\delta_{nm}$& 
Geometrical& $g^{2}(0)=2$\\ \hline
$P(\alpha)=\delta^{2}(\alpha-\alpha')$&$\rho_{nn}=\frac{|\alpha'|^{2n}
e^{-|\alpha'|^{2}}}{\Gamma(n+1)}$&Poissonian&$g^{2}(0)=1$\\ \hline
$P(\alpha)=\frac{e^{|\alpha-\alpha'|^{2}/\overline{n}}}{\pi\overline{n}}$&
$\rho_{nn}=\frac{\overline{n}^{n}}{(1+\overline{n})^{n+1}}
e^{-|\alpha '|^{2}/(1+\overline{n})}$& Glauber-Lachs&
$1\le g^{2}(0)\le2.$\\
&$ \times L_{n}(\frac{-|\alpha'|^{2}}{\overline{n}(1+\overline{n})})$&&\\ \hline
K sources& & &       \\  \hline
Gaussian (Chaotic)&$\rho_{nn}=\frac{n+k-1)!}{n!(k+1)!}(\frac{\overline{n}/k}
{1+\overline{n}/k})^{n}\frac{1}{(1+\overline{n}/k)^{k}}$&Negative Binomial
&""\\ \hline
Coherent +Chaotic&$\rho_{nn}= \frac{(p\overline{n}/k)^{n}}
{(1+p\overline{n}/k)^{n+k}}$& Perina-McGill& """"\\
&$\times e^{\left[ \frac{-\gamma p\overline{n}}{1+p\overline{n}/k}\right]}
L_{n}^{k-1}(\frac{-\gamma k}{(1+p\overline{n}/k)})$&&\\ \hline
\end{tabular}    \\
\end{table}

\section{The Phenomenological model}
Experimentally there exists a large class of data ($\nu p$) and low mass 
diffractive data that have mulplicity distributions  with sub-Poissonian 
Statistics.
Thus we  seek a more general distribution than the ones given in table 1.
A clue as to the  appropriate distribution is that charged pions occur in pairs
 Furthermore the most general Gaussian source characterised by Gaussian 
 Wigner Function . These facts point to the use of  Squeezed Coherent 
 states.\\
 We find that the  k-mode squeezed state
$|\alpha,r>=|\alpha_{1},r_{1}>|\alpha_{2},r_{2}>\cdots |\alpha_{k},r_{k}>$\\
characterised by the
multiplicity distribution:
\begin{equation}P_{n}{}^{k}=\Pi_{i}^{k}P_{n_{i}} \cdots \sum_{i}n_{i}=n
\end{equation}
\begin{eqnarray}
P_{n}{}^{k}&=&e^{\left[ -k\alpha^{2}(1+x)\right]} (1-x^{2})^{k/2}
(\frac{x}{2})^{n} \nonumber \\
&\times&\sum_{m=0}^{\left[\frac{n}{2}\right]}\frac{\gamma_{m}H_{n-2m}{}^{2}
(\sqrt{k}y)2^{2m}}{m! (n-2m)!} \nonumber
\end{eqnarray}
$y=(\frac{\alpha^{2}(1+x)^{2}}{2x})^{\frac{1}{2}}$\\
 $\gamma=\frac{k-1}{2}$; $\gamma_{m}=(\gamma+1)\cdots (\gamma+(m-1))\;\; ; 
 \; \gamma_{0}=1$\\
and the
second order correlation function:
\begin{equation}g^{2}_{k}(0)=1+ \frac{2sinh^{4}(r)+(2\alpha^{2}+1)sinh^{2}(r)
-sinh(2r)}{k(\alpha^{2}+sinh^{2}(r))^{2}}\end{equation}
 is the most general distribution that fits a wide range of data \cite{bam}.
If $r>0$ there are regions where $g^{2}_{k}(0) <1 $  and the distribution is 
narrower than Poissonian.If $r<0$; $g^{2}_{k}(0)$ is always greater then 1 
showing distributions which are broader than Poissonian.       

Hadronic distributions in $p\overline{p}$ collisions show broader 
than Poisonnian multiplicity distributions with a long multiplicity tail,
which gets broader and broader with the increase of energy. The $k=3$ mode
distribution for $\overline{n}=13.6$, $x=-0.20$ and $\overline{n}=26.1$, 
$x=-0.35$ respectively fit corresponding ISR ($62.2 Gev$) and UA5 ($540 Gev$) 
data, $\alpha$ for each of these is thus fixed.\\
To fit neutrino induced collisions in which the distribution is 
super-Poissonian ($(\frac{\Delta n}{\overline{n}})<1$),$k=3, x=0.5$ fit  
data well. $e^{+}e^{-}$ collisions are fit by the k=2 squeezed coherent 
distribution with r close to zero . (nearly Poissonian.)

\section{The Statistics confronts the Dynamics}
 We would now like to conjecture on the reason for this success and find an
overlap with dyamical models.
We search for incoming states of the hadronic fireball which will give rise 
to SQUEEZED COHERENT DISTRIBUTION.
 The candidate dynamical model of hadrons, which we find is appropriate is the
covariant phase space model for hadrons  which is a
 revival of Feynman et. al's relativistic harmonic oscillator model\cite{fey}
Kim and Wigner pointed out that the covariant harmonic
 oscillator model is the natural language
for a covariant description of phase-space \cite{kim89}, \cite{kim90}.
In this paper, we use the covariant phase space distribution description of
relativistic extended particles to give a phenomenological description of 
multiplicitydistributions in the high energy collisions of hadrons

Wave functions without time-like oscillations can be constructed by
using the unitary representations of the Poincare group and
imposing  a covariant
condition\cite{kim86}.  In this model
 two quarks bound together by a relativistic harmonic oscillator potential
mapped onto
O(3,1) invariant harmonic oscillator equation.
The ground state wave function $\psi^{0}_{\beta}$ in the Lorentz boosted 
(primed) frame is
\begin{equation} 
\psi_{0}^{\beta}(x')=[e^{-i\eta K_{3}}]\psi_{0}(x)
\end{equation}
where $K_{3}$ boost generator along the z axis,
\begin{equation} 
K_{3}=i(z\frac{\partial}{\partial t} + t\frac{\partial}{\partial z}) 
\end{equation}
 and $ \eta=Tanh^{-1}(\beta)$

 In `Quantum Optical' language,  using light-
cone variables we have:
\begin{equation}
u=(t+z)/\sqrt{2} \; \; \; v=(t-z)/\sqrt{2} \nonumber
\end{equation}
Then in the Lorentz-transformed frame;
\begin{eqnarray}
 q'_{u}=e^{\eta}q_{u}  \;\;\; q'_{v}=e^{-\eta}q_{v} \nonumber \\
u'=e^{-\eta}u  \;\;\; v'=e^{\eta}v
\end{eqnarray}
Introducing creation and annihilation operators
\begin{eqnarray}
a_{u}=u+\frac{\partial}{\partial u}.... 
a_{v}=v+\frac{\partial}{\partial v}\nonumber \\
a^{\dagger}_{u}=u-\frac{\partial}{\partial u}......
a^{\dagger}_{v}=v-\frac{\partial}{\partial v} \nonumber
\end{eqnarray}
 we find that the wave function $\psi_{n}^{\beta}=\psi(u',v')$ is a two mode 
 squeezed state.
\begin{equation}
\psi_{0}^{\beta}(u',v')=|0,\beta> =|0,\eta>_{u'}|0,-\eta>_{v'}
\end{equation}
The excited state is given by:
\begin{equation}
|n,\beta>=(a_{u'}^{\dagger})^{n}|0,\eta>(a_{v'}^{\dagger})^{n}|0,-\eta> 
\end{equation}
The condition for  absence of time-like oscillations in the hadronic 
rest-frame 
\begin{equation}
(a'_{u} -a'_{v})|n,\beta>=0.
\end{equation}
 The physical wave functions are
\begin{equation} 
\psi_{n}^{\beta}(u',v')=|n,\beta> =\sum_{m=0}^{n} \left(\begin{array}{c}n\\m 
\end{array}\right) |n-m,\eta>_{u'}|m,-\eta>_{v'} 
\end{equation}
in the Fock-space representation
\begin{equation}
\psi_{n}^{\beta}(u',v')=\sum_{n_{1},n_{2}}\sum_{m=0}^{n}
\left(\begin{array}{c}n\\m \end{array}
\right)G_{n_{1},n-m}(\eta)G_{n_{2},m}(-\eta)
\psi_{n_{1}}^{0}(u)\psi_{n{2}}^{0}(v)   
\end{equation}
Where \cite{sat}:
\begin{eqnarray}
G_{n,m}=
(-1)^{\frac{m+n}{2}}(\frac{m! n!}{cosh(\eta)})(\frac{tanh(\eta)}{2})^
{\frac{m+n}{2}} \nonumber\\
\times \sum_{\lambda}^{min[\frac{n}{2}\frac{m}{2}]}
\frac{(\frac{-4}{sinh^{2}(\eta)})^{\lambda}}{(2\lambda)!(m/2-\lambda)!
(n/2-\lambda)!}
\end{eqnarray}
for n,m even
and
\begin{eqnarray}
G_{n,m}=(-1)^{\frac{m+n}{2}-3/2}(\frac{m! n!}{cosh(\eta)})
(\frac{tanh(\eta)}{2})^{\frac{m+n}{2}-1} \nonumber \\
\times \sum_{\lambda}^{min[\frac{n-1}{2},\frac{m-1}{2}]}
\frac{(\frac{-4}{sinh^{2}(\eta)})^{\lambda}}{(2\lambda+1)!(m-1/2-\lambda)!
(n-1/2-\lambda)!} 
\end{eqnarray}
for n,m odd.
 $ G_{n,m}$ is non zero for both n,m even or both n,m odd , thus excitations of 
 quarks occur in pairs.
and the
 Lorentz squeezed vacuum is a many particle state .
 The above suggests the identification of Hadronic sources in terms of squeezed 
 states.\\
In the `fireball picture ' the
Wigner function  of the source is , \cite{cha}
\begin{eqnarray}
W^{n}(u,v,q_{u},q_{v})&=&
(\frac{2}{\pi})^{2}e^{-\frac{1}{2}(e^{\eta}u^{2}+e^{-\eta}q_{u}^{2} 
+e^{-\eta}v^{2}+e^{\eta}q_{v}^{2})} \nonumber \\
&\times&\sum_{m=0}^{n}\left( \begin{array}{c}n\\m \end{array} \right)
(-1)^{n} L_{n-m}[e^{\eta}u^{2}-e^{-\eta}q_{u}^{2}]
L_{m}[e^{-\eta}v^{2}-e^{\eta}q_{v}^{2}]
\end{eqnarray}
 The number
 distribution for l particles in the $n^{th}$ excited state.
\begin{equation} 
P_{l}=\sum_{l_{1}+l_{2}=l}\sum_{m=0}^{n}\left(\begin{array}{c}n\\m\end{array}
\right)P_{l_{1}}^{n-m,sq}(\eta)P_{l_{2}}^{m,sq}(-\eta)
\end{equation}
\begin{eqnarray}
P_{l_{2}}^{m,sq}&=&
\frac{(m)!l_{2}!}{(cosh(-\eta))^{2l_{2}+1}}(\frac{tanh(-\eta)}{2})^{m-l_{2}} 
\nonumber \\
&\times&F(-\eta,l_{2},m)\cos^{2}(\frac{(m-l_{2})\pi}{2}) 
\end{eqnarray}
where:
\begin{equation}
F(-\eta,l_{2},m)= \sum_{\lambda=0}^{min(l_{2}/2,(m-l_{2})/2}
\frac{(\frac{-4}{sinh^{2}(-\eta)})^{\lambda}}{(\lambda)!(l_{2}-2\lambda)!
(m-l_{2}-2\lambda)!}
\end{equation}
The cosine terms imply $P_{l}$ vanishes when $|m-l_{2}|$ or $|n-m-l_{1}|$ is odd. 
So that the excited each of
oscillator modes is excited in pairs. \\
{\bf If each pair is associated with a two quark bound state(pion), the excited 
state contains pair correlated
pions!!!}
\section{Results and Conclusion}
The picture emerging is as follows
 the distribution of the fireball results from
the excitation of oscillator modes of the colliding hadrons.
This excitation takes place
in pairs.
Modes de-excite statistically emitting 2 pairs of quarks which we identify
as two pions
 The phase space
distribution of the fireball:
\begin{equation}
|<n,\beta|n,-\beta'>|^{2}=
(2\pi)\int dudvW^{n}_{\beta}(u,v,q_{u},q_{v}) 
W^{n}_{\beta'}(u,v,q_{u},q_{v}) 
\end{equation}
Probability of emission of m particles from
two independent populations 1 and 2 corresponding to each of the incident 
hadrons.
forming an overlapping distributions is given as:
\begin{equation}P_{m}=\sum_{m'=0}^{m}P^{1}_{m-m'}P^{2}_{m'} \end{equation}
 Total probability distribution thus becomes a product of
 the probability distribution of four squeezed sources:
\begin{eqnarray}
P_{m}=\sum_{{m_1}+-m_{2}+m_{3}+m_{4}=m}P_{m_{1}}^{sq}(\eta)P_{m_{2}}^{sq}
(-\eta) 
P_{m_{3}}(\eta')P_{m_{4}}(-\eta') \nonumber
\end{eqnarray}

 For target Projectile collisions  $\beta'=0$  thus the
probability of emitting n' particles is:
\begin{eqnarray}
P_{n'}&=&\sum_{p=0}^{n'}\sum_{m=0}^{n}\left( \begin{array}{c}n'\\p\end{array}
\right)
\left(\begin{array}{c}n\\m \end{array} \right) \nonumber \\
&\times &(-1)^{\frac{n'+n}{2}}(p! (n-m)!m!(n'-p)!)^{1/2}\frac{1}{cosh(\eta)}
(\frac{tanh(\eta)}{2})^{\frac{n'+n}{2}}
(-1)^{\frac{m+n'-p}{2}} \nonumber \\
&\times & \sum_{\mu\lambda}\frac{(\frac{-4}{sinh^{2}(\eta)})^{\lambda + \mu}}
{(2\mu)!(p/2-\mu)!(\frac{n'-p}{2}-\mu)!(m/2-\lambda)!(\frac{n-m}{2}-\lambda)!} 
\nonumber
\end{eqnarray}
As $\beta$ increases the distribution gets broader .

For Central Collisions $\beta=\beta'$ and by plotting
$mP_{m}$ vs.$\frac{m}{<m>}$for different values of $\beta$ we see that
the distributions become
wider and skew symmetric as the value of $\beta$ becomes larger. This is
consistent with the variation seen in experimental data.

The total probability
distribution for the two nucleon system for n pions is:
\begin{equation}P_{n}^{k}=\sum_{\sum n_{i}=n} \prod_{i}^{k/2}P_{n_{i}}^{sq}
(\eta)
\sum_{\sum n_{i}=n}\prod_{i}^{k/2}P_{n_{i}}^{sq}(-\eta) \end{equation}
where k=6 for nucleon-nucleon collisions ,
k=4 for $\pi\pi$ collisions and  k=3 for $\nu p$ collisions (with $\eta$ 
positive).

We include final state interactions  in a simple fashion  by assuming that
the effect of interaction is to add coherence into the final state. This is  
consistent
with the fact that in particle collisions experimental data shows some amount of 
coherence, especially
in the low energy region , among the emitted particles.
With the resulting density matrix we obtain the mutiplicity distribution for a 
variety of collisions and compare to data.
The distribution we get is:
\begin{equation}
P_{n}^{k}=\sum_{\sum n_{i}=n} \prod_{i}^{k/2}P_{n_{i}}^{sq, coherent}
(\alpha,\eta)
\sum_{\sum n_{i}=n}\prod_{i}^{k/2}P_{n_{i}}^{sq , coherent}(\alpha,-\eta)
\end{equation}
Where the average number of particles emitted by each mode is given by
:$\overline n_{i}=\alpha^{2}+Sinh^{2}(\eta)$
Above distribution fits the CERN ISR 62.2 GeV and UA5 540 GeV data.
 The k=3 distribution is compared with $\nu p$ data.
The data is well reproduced by the distribution.
For $e^{+}e^{-}$ collisions we take k=2 because the intermedeate  
state is the virtual $\overline qq$ state formed by the colliding electron and 
positron. \\
  
In terms of hadronic final states
the LEP energy ($\sqrt{s}=100$ GeV) is equivalent to the SPS energy 
($ \sqrt{s}=546$ GeV )  as far as total mutiplicities are concerned, in so far 
as $\overline n^{e^{+}e^{-}} 
$(LEP) $\approx\overline n^{\overline p p} $(SPS) $\approx$26 .\\
For the same value of $\overline{n}$ much narrower distribution
for $e^{+}e^{-}$ distributions than  the $\overline{p} p $ distributions.
This is consistent with recent LEP data \cite{Al}.

 We can make some predictions for higher energies such as those observed 
 at the LHC and SSC.
Since widening of the distributions is related to the squeezing parameter 
$\eta$ 
the lorentz boost of the hadronic fireball,
at
C.M.S. energies of 20 TeV and above  we have a large $\beta$ value and
higher modes will be excited.
 The multiplicity distribution for ultra-high energies is  very broad and 
 skew-symmetric.
plot $\overline{n}P_{n} vs.\frac{n}{\overline{n}}$ for $\overline{p}p$ collisions 
for $\overline  n=50$

We can also calculate the
Bose-Einstein Correlations of pions in this model by using the two mode state.
Ongoing work is in progress to establish the connection of this model with QCD 
using the light cone formalism \cite{bro}.
In this formalism it is also easy to incorporate temperature dependence by 
using Thermal Squeezed Coherent states. These would be of interest in heavy 
ion collisions.
\section{Acknowledgements}
I would like to thank Prof. Y.S. Kim and all the other organizers of 
ICSSUR '95 for their kind hospitality and support. I also acknowledge the 
support of the University Grants Commision,National Board of Higher Mathematics, 
INSA , DST and CSIR for their support.

\end{document}